\documentclass[]{spie}  
\usepackage[]{graphicx}
\usepackage{epstopdf}
\usepackage{bpchem}
\usepackage{graphicx, subfigure}
\usepackage[mediumspace,mediumqspace,Grey,squaren]{SIunits}

\title{Terahertz Coded Aperture Mask using a Vanadium Dioxide Bowtie Antenna Array } 

\author{Souheil Nadri\supit{a}, Rebecca Percy\supit{a}, Lin Kittiwatanakul\supit{b}, Alex Arsenovic\supit{c}, Jiwei Lu\supit{b}, Stu Wolf\supit{b}, Robert M. Weikle II\supit{a} 
\skiplinehalf
\supit{a}University of Virginia, Department of Electrical and Computer Engineering, Charlottesville, VA 22904 U.S.A;  \\
\supit{b}University of Virginia, Department of Materials Science and Engineering, Charlottesville, VA 22904 U.S.A;  \\
\supit{c}Virginia Diodes, Charlottesville, VA 22904 U.S.A
}

\begin{document} 
  \maketitle 

\begin{abstract}
Terahertz imaging systems have received substantial attention from the scientific community for their use in astronomy, spectroscopy, plasma diagnostics and security. One approach to designing such systems is to use focal plane arrays. Although the principle of these systems is straightforward, realizing practical architectures has proven deceptively difficult. A different approach to imaging consists of spatially encoding the incoming flux of electromagnetic energy prior to detection using a reconfigurable mask. This technique is referred to as \textquotedblleft{coded aperture}\textquotedblright or \textquotedblleft{Hadamard}\textquotedblright imaging. This paper details the design, fabrication and testing of a prototype coded aperture mask operating at WR-1.5 (500-750 GHz) that uses the switching properties of vanadium dioxide(VO\textsubscript{2}). The reconfigurable mask consists of bowtie antennas with vanadium dioxide VO\textsubscript{2} elements at the feed points. From the symmetry, a unit cell of the array can be represented by an equivalent waveguide whose dimensions limit the maximum operating frequency. In this design, the cutoff frequency of the unit cell is 640 GHz. The VO\textsubscript{2} devices are grown using reactive-biased target ion beam deposition. A reflection coefficient ($S_{11}$) measurement of the mask in the WR-1.5 (500-750 GHz) band is conducted. The results are compared with circuit models and found to be in good agreement. A simulation of the transmission response of the mask is conducted and shows a transmission modulation of up to 28 dB. This project is a first step towards the development of a full coded aperture imaging system operating at WR-1.5 with VO\textsubscript{2} as the mask switching element. 
\end{abstract}


\section{INTRODUCTION}
\label{sec:intro}  

Millimeter-wave imaging arrays have been an important topic of research for the scientific community for decades and they continue to receive significant attention for their use in a variety of important applications, including radio astronomy, imaging spectroscopy, and plasma diagnostics. Moreover, imaging at terahertz or submillimeter-wave frequencies has begun to attract enormous interest because of its large potential impact on defense, security, the biological sciences, medicine and materials science [1-2]. Millimeter-wave imaging systems generally fall into one of two fundamentally different categories: scanned systems and focal plane arrays. The use of the former method is limited to applications in which a scene or object being observed does not change rapidly. Moreover, it exhibits a limited signal-to-noise ratio due to practical integration times. The latter method consists of an ensemble of millimeter-wave sensors and a focusing element. Although the principle of focal plane imaging is straightforward, realizing practical architectures for these systems has proven difficult. A primary issue is “floor-planning” (accommodating the output signals from each pixel). The difficulty lies in designing output feeds for the various elements in the array without significantly disturbing its operation. The challenge of accommodating multiple IF outputs usually precludes the use of heterodyne detection in such arrays. The system requirements of imaging arrays for many present-day applications, however, place such stringent requirements on performance that they either eliminate consideration of direct detection schemes (due to their low sensitivity compared to heterodyne detection), or push the technology to its fundamental limits (greatly increasing cost and complexity).

An alternative approach to imaging based on the focal plane technique described above is to spatially encode the incoming flux of electromagnetic energy prior to detection. This technique is generally referred to as \textquotedblleft{coded aperture}\textquotedblright or \textquotedblleft{Hadamard}\textquotedblright imaging and it permits multiple image points to be superimposed (or multiplexed) onto a single detector. This technique offers unique advantages. Foremost among them is that spatial encoding and multiplexing can be used to eliminate the need for an array of detectors. Thus, the floor-planning issue, which is inherent to large detector arrays, is eliminated. A second important advantage of the coded aperture approach is that it more readily accommodates heterodyne detection methods. Furthermore, Hadammard imaging offers a significant improvement in signal-to-noise ratio (SNR) compared to comparable scanned imaging system [3].

The central component of a coded aperture imaging system is the mask that modulates the incident wavefront. Traditionally, this mask has consisted of opaque and transparent apertures. Different mask patterns are obtained by moving the mask mechanically, resulting in different unique patterns. In this paper, we report on a vanadium dioxide bowtie array mask in which the mask elements are switched thermally. The array is found to be capable of modulating the transmission of a 500 to 750 GHz signal by up to 28 dB. A 100 nm thick VO\textsubscript{2} film located at the antenna feed point modulates the reflectance of the bowtie antenna such that it reflects the submillimeter wave signal when the VO\textsubscript{2} is in the “off” state (semiconducting) and transmits the signal when it is in the “on” state (metallic) [4]. A WR-1.5 one-port measurement system was assembled to validate the model of the bowtie antenna array. From measurements, the properties of the sapphire substrate and the VO\textsubscript{2} film are extracted. The transmission response of the mask is simulated and its performance predicted. In the final section, we investigate another method of switching the array using laser illumination.

\section{Vanadium Dioxide} 

\subsection{Background} 
\label{sec:title}

Vanadium dioxide is a Mott insulator, so that due to electron-electron interactions, a bandgap of 0.7 eV is formed which prevents conduction of electrons at low temperatures. However, VO\textsubscript{2} is a material that exhibits an abrupt and reversible semiconductor-to-metal transition (SMT) at a temperature of 67\degree{}C, close to room temperature. This proximity of the transition to room temperature makes VO\textsubscript{2} appealing for many applications. Moreover, this transition can be triggered through current injection, application of an electric field [6] or optical pumping [7]. When VO\textsubscript{2} undergoes a change from semiconductor to metal, it undergoes large modifications in its electrical and optical properties. The SMT is accompanied by changes in resistivity and reflectivity of the film. The change in resistivity varies from about one-half order of magnitude to five orders of magnitude depending on the transition method and film parameters [7]. A change in transmission of more than $40\%$ is reported between the semiconducting and metallic state at terahertz, infrared, and optical frequencies [4-7]. These properties make VO\textsubscript{2} attractive for a terahertz switch capable of modulating a signal beam. 

\subsection{Fabrication} 
The VO\textsubscript{2} film used in this work was grown at the University of Virginia NanoSTAR Institute Department of Materials Science using reactive-biased target ion beam deposition (RBTIBD). This technique is used to deposit the VO\textsubscript{2} thin film onto a c-plane sapphire substrate. The main reactor chamber is pumped to a base pressure of $5*10^{-8}$ Torr, and the stage heater is set to 400-$500\,^{\circ}\mathrm{C}$ and allowed to stabilize for 45 minutes. The substrate surface is sputter cleaned using Ar ions and the vanadium target is sputter-cleaned by applying pulsed dc bias. Vanadium was sputtered in an 80/20 mixture of Ar and O\textsubscript{2} at a flow rate of 5.0 to 6.0 sccm, which results in a pressure of approximately 1 mTorr. A deposition time of approximately 2 hours results in a film thickness of about \unit{5}{\nano\meter}.

\section{Array Architecture} \label{sec:sections}

The switching array consists of a 20 by 20 square matrix of  bowtie antennas. The bowtie antennas have an angle of $90^{\circ}$, a diagonal length of \unit{71.7}{\micro\meter} (figure 1(a)) and are made of a \unit{0.15}{\micro\meter} thick layer of gold. Gold is also deposited on either side of the array to serve as electrical contacts seen in figure 1(a). These contacts are used to measure the resistance of the array versus temperature and the Mott transition, which is characterized by a change in resistance of two orders of magnitude and a hysteretic behavior [8] (figure 2). 
\begin{figure}[ht!]
     \begin{center}
        \subfigure[]{%
            \label{fig:first}
            \includegraphics[width=0.6\textwidth]{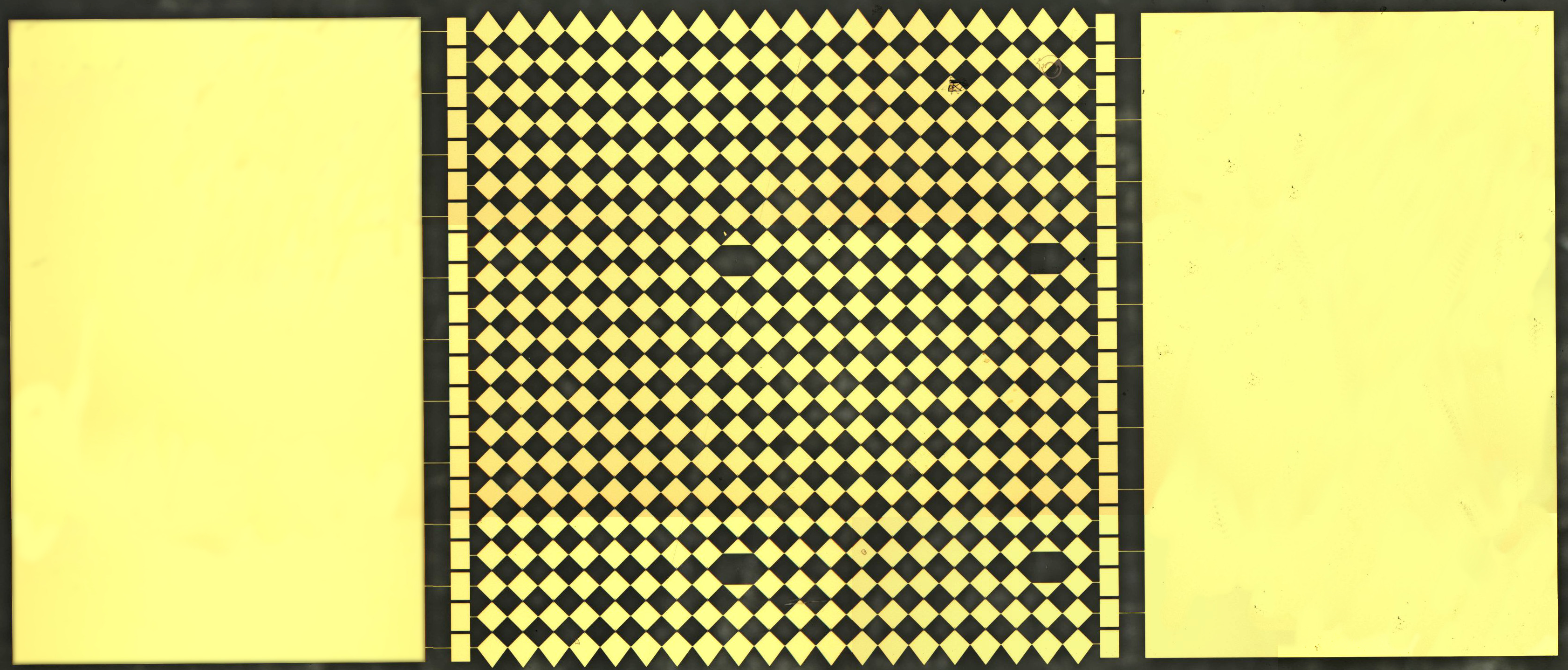}
        }%
        \subfigure[]{%
           \label{fig:second}
           \includegraphics[width=0.4\textwidth]{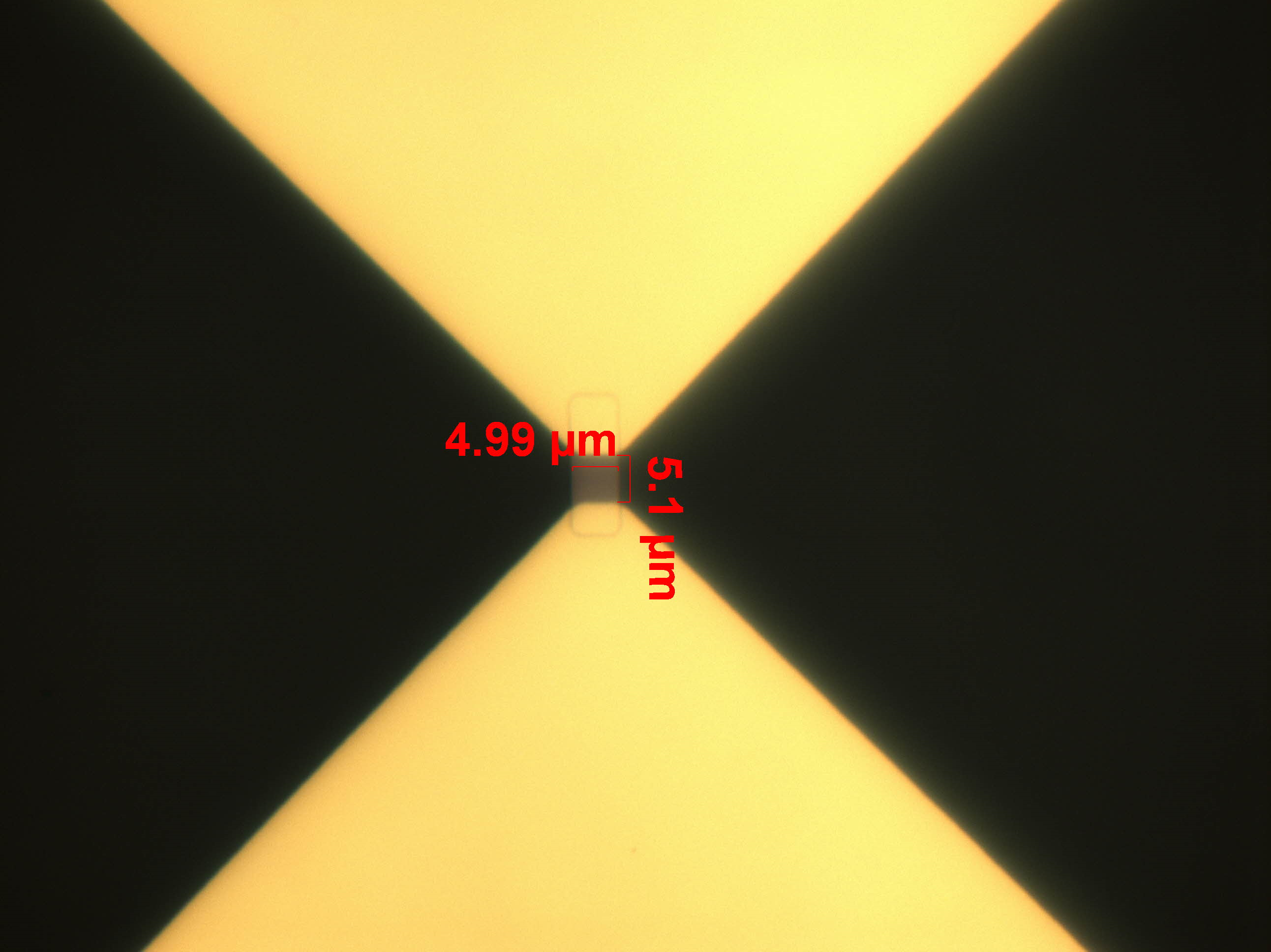}
        }\\ 
        
    \end{center}
    \caption{
        (a)\hspace{1 mm}Microscope image of the bowtie antenna array. The rows are connected in parallel. The figure also shows the two gold contact pads on both sides. Some elements are missing due to defects in the mask. The structure is built on a sapphire substrate.(b)\hspace{1 mm}Higher magnification image of a single bowtie. The vanadium dioxide film is located at the antenna feed.
     } 
     \label{fig:subfigures}
\end{figure}
\begin{figure}[h]
   \begin{center}
   \begin{tabular}{c}
   \includegraphics[height=6cm]{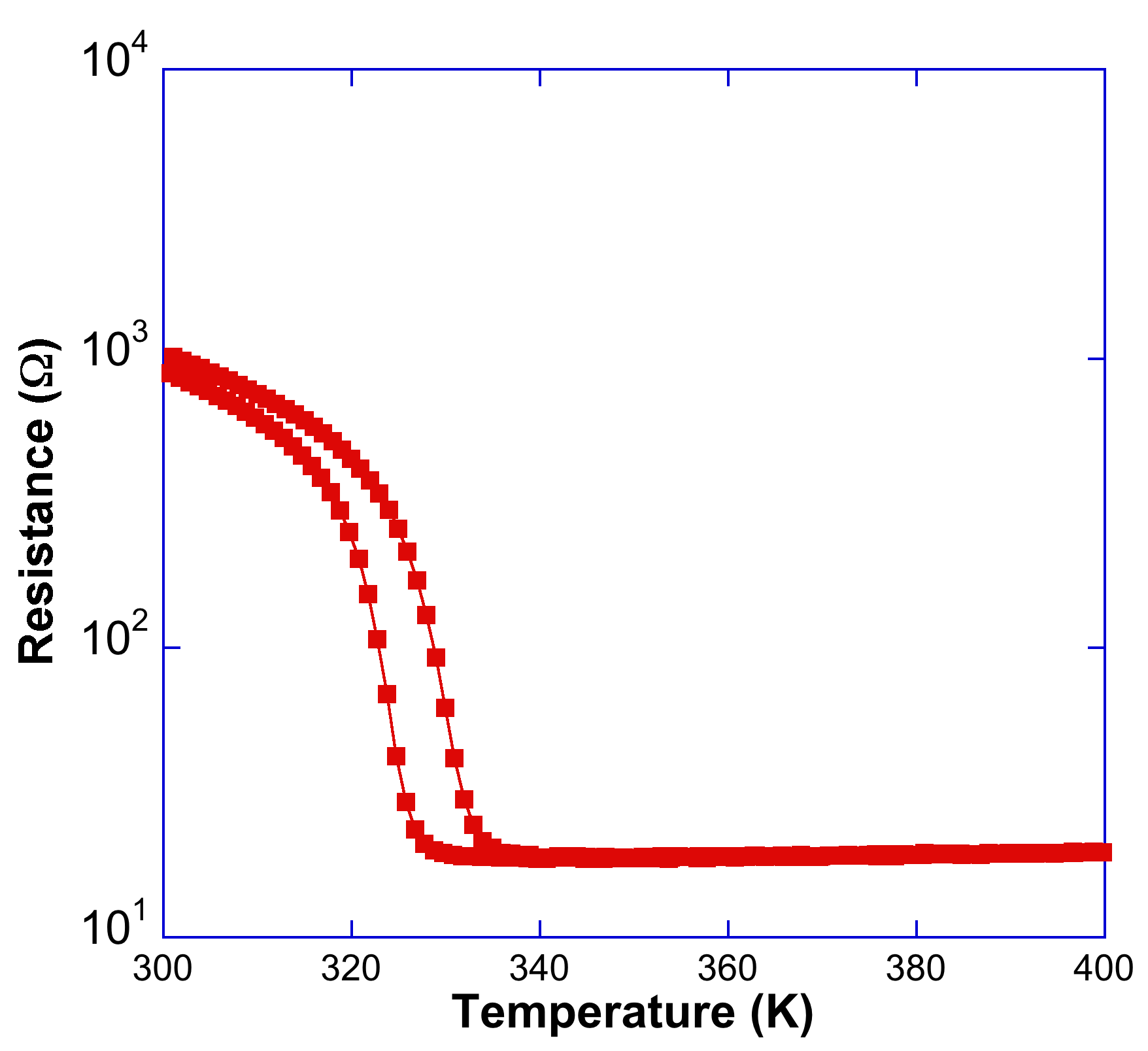}
   \end{tabular}
   \end{center}
   \caption[example] 
   { \label{fig:example} 
Resistance vs temperature measurement of the sample. The measurement is done in a chamber that allows very accurate control of the temperature. The resistance changes by about two orders of magnitude and the behavior is hysteretic. }
   \end{figure}
A VO\textsubscript{2} bridge is located at the antenna feed and has a size of \unit{5}{\micro\meter} by \unit{5}{\micro\meter} (figure 1(b)). \\
From symmetry, the array unit cell can be represented as an equivalent waveguide with virtual electric walls along horizontal symmetry planes and virtual magnetic walls along vertical symmetry planes. This equivalent waveguide is represented by the unit cell circuit model shown in figure 3(a) [9], and has a size of $\unit{69}{\micro\meter}$ by $\unit{69}{\micro\meter}$. These dimensions were chosen to only allow TEM mode propagation below 640 GHz. Above this cutoff frequency, the performance of the array deteriorates due to multimode propagation and formation of grating lobes. From the equivalent waveguide model, we can construct a circuit model representation for the array. The bowtie structure itself can be represented as an equivalent transmission line with characteristic impedance of Z\textsubscript{bow} and an electrical length of $\theta\textsubscript{bow}$. The values for these parameters can be found using the EMF method [10] or finite-element electromagnetic simulation with HFSS. Furthermore, the vanadium dioxide bridge is modeled as a shunt temperature-dependent variable resistor. Finally, the air and sapphire substrates are also modeled as transmission lines. A representation of the array circuit model is shown in figure 3(b).

\begin{figure}[ht!]
     \begin{center}
        \subfigure[]{%
            \label{fig:first}
            \includegraphics[width=0.4\textwidth]{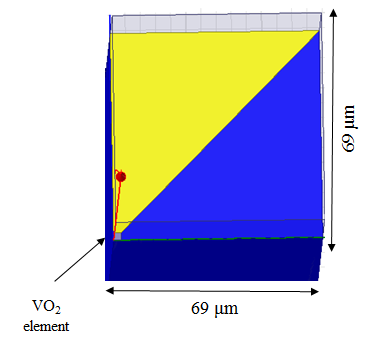}
        }%
        \subfigure[]{%
           \label{fig:second}
           \includegraphics[width=0.5\textwidth]{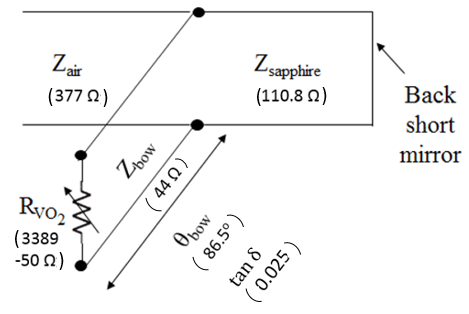}
        }\\ 
        
    \end{center}
    \caption{
        (a)\hspace{1 mm}Layout of the array unit cell represented as an equivalent waveguide. The top and bottom walls are perfect E-walls. The right and left walls are perfect H-walls. The sapphire (in blue) has a thickness of \unit{400}{\micro\meter}. The vanadium dioxide film lies at the bottom left corner of the unit cell. Both the gold layer (in yellow) and the VO\textsubscript{2} have a thickness of \unit{.15}{\micro\meter}. (b)\hspace{1 mm} Circuit model representing the bowtie antenna array. The vanadium film is represented as a variable resistor, with value ranging from $3.4\hspace{1 mm} k\Omega$ to $50\hspace{1 mm} k\Omega$ dependent on temperature.
     } 
     \label{fig:subfigures}
\end{figure}  
 
\section{Reflection measurements} \label{sec:sections}
\subsection{Measurement Apparatus}
A reflection coefficient measurement setup to characterize the array is shown in figure 4. A Rohde \& Schwarz (ZVA-40) vector network analyzer together with a VDI WR-1.5 extension module provide the incident 500-750 GHz radiation through a diagonal horn antenna. The beam is collimated by an initial off-axis parabolic mirror (focal length of 76.4 mm). A second parabolic mirror (f=76.4 mm) focuses the beam onto the sample (placed on an electronic hot plate) using a 45\degree{} mirror. The reflected beam travels back through the same path to be detected by the VNA. For the reflection coefficient($S_{11}$) measurement, a back short mirror is placed behind the device under test (DUT) to eliminate possible absorption from the material of the hot plate surface.

\begin{figure}[h]
   \begin{center}
   \begin{tabular}{c}
   \includegraphics[height=6cm]{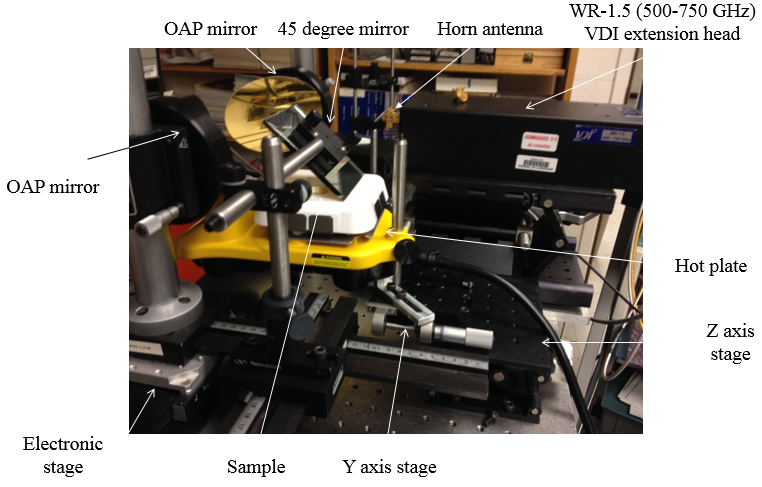}
   \end{tabular}
   \end{center}
   \caption[example] 
   { \label{fig:example} 
Reflection coefficient($S_{11}$) measurement setup for the WR-1.5 frequency range. The mirrors are used to direct the beam onto the sample that sits on the electronic hot plate. We can therefore ensure precise control and reading of the temperature of the hot plate and sample.}
   \end{figure}    
 
\subsection{Calibration} 
Before useful measurements can be made with the quasi-optical setup system, the system must be calibrated by measuring a set of known standards. Calibration is used to remove the unwanted systematic errors associated with the measurement system [11]. Five delay shorts (with delays increments of \unit{40}{\micro\meter}) and a low reflection load (microwave absorber) are used as standards. The second parabolic mirror, along with the 45\degree{} mirror and the hot plate, are mounted to an electronic translation stage. By moving the stage, the phase delay between the measurement reference plane and reflect standard can be adjusted in an accurate and repeatable way. In addition, as more standards than necessary are used, the calibration is overdetermined. As a result, an improvement in estimation of the error coefficients of the system can be achieved compared to using a minimum set of three [11]. All the algorithms and calculations used to correct the raw data have been made with the open-source python module scikit-rf [11]. A calibrated measurement of the one-port calibration standards is shown in figure 5.

\begin{figure}[h]
   \begin{center}
   \begin{tabular}{c}
   \includegraphics[height=8cm]{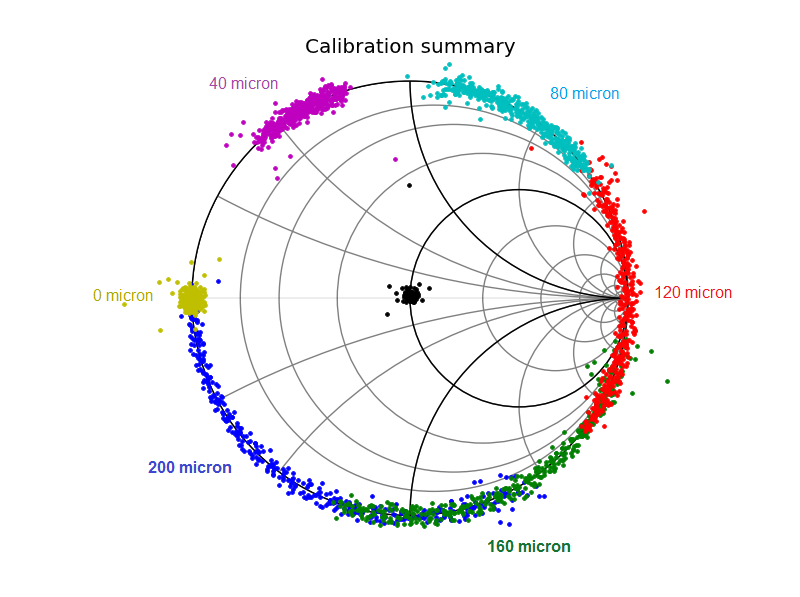}
   \end{tabular}
   \end{center}
   \caption[example] 
   { \label{fig:example} 
Measurement of the system calibration standards from 500-750 GHz shown on a Smith chart. Microwave absorber is used as a match, at the center of the smith chart. The delay shorts are on the Smith chart perimeter and do not overlap, as required for a well-behaved calibration.}
   \end{figure}  
To verify the measurements, a sapphire(Al\textsubscript{2}O\textsubscript{3})substrate (400 {\micro\meter} thick) with backshort was measured as a verification standard. As seen in figure 6(a), sapphire is lossless over the WR-1.5 band and therefore, all the power is reflected back to the VNA. Note that the magnitude of the reflection coefficient in figure 6(a) wanders above zero at some frequencies which is due to uncorrected measurement error. Furthermore, by fitting the phase measurement to the transmission line model, shown in figure 6(b), we find that the reference plane is shifted by 40 {\micro\meter} (due to a misalignment error) and that the dielectric constant of sapphire is 11.58 parallel to the c-axis (in agreement with its nominal value). The model also includes an air gap prior to the reference plane (due to a slight change of thickness between the calibration standards and the DUT), and an air gap between the sample and the back short.  
\begin{figure}[ht!]
     \begin{center}
        \subfigure[]{%
            \label{fig:first}
            \includegraphics[width=0.5\textwidth]{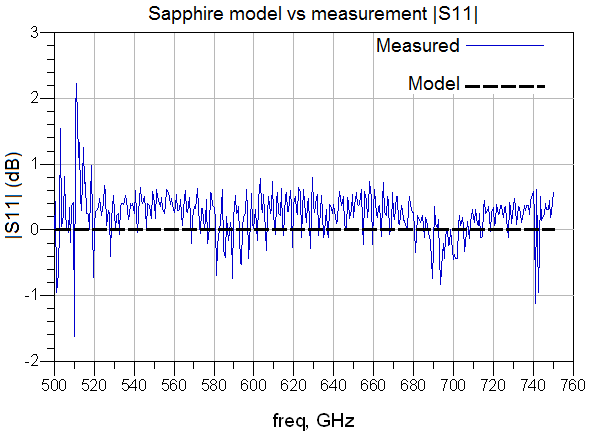}
        }%
        \subfigure[]{%
           \label{fig:second}
           \includegraphics[width=0.5\textwidth]{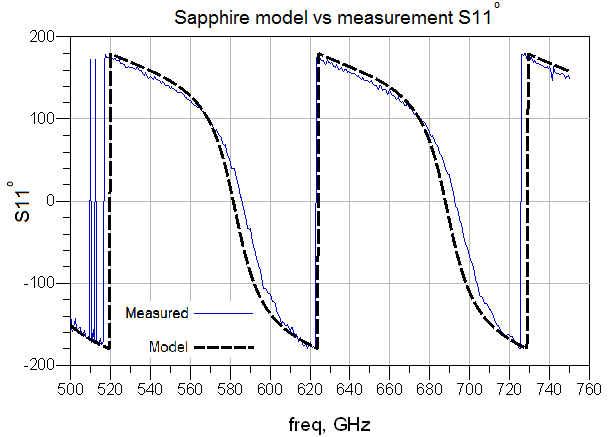}
        }\\ 
        
    \end{center}
    \caption{
        (a)\hspace{1 mm} Magnitude of $S_{11}$ for a sapphire substrate with backshort conductor. Sapphire is lossless over the WR-1.5 band and therefore all the energy is reflected back. There is a deviation by $^+_-1$ dB due to calibration error. (b)\hspace{1 mm} Phase of $S_{11}$ (in degrees) for a piece of sapphire. The thickness of the piece of sapphire is 400 {\micro\meter}. The model also includes an air gap prior to the reference plane, and an air gap between the sample and the back short.  
     } 
     \label{fig:subfigures}
\end{figure}  
\newpage
\subsection{Beam size} 
The size of the VO\textsubscript{2} mask array, excluding the contact pads is 7.2\hspace{1 mm}mm by 7.2\hspace{1 mm}mm. An experiment was set up to determine the beam spot size, which should be no larger than the size of the array being measured. To ensure that this requirement is met, we cut a thin rectangular piece  of microwave absorber the size of the array and placed it on a back short mirror on the hot plate. The response shown in Figure 7 shows little reflection which verifies that the beam is sufficiently focused for this measurement. Note that the reflection peak at 515 GHz is likely associated with error uncorrected by the calibration. 
\begin{figure}[h]
   \begin{center}
   \begin{tabular}{c}
   \includegraphics[height=6cm]{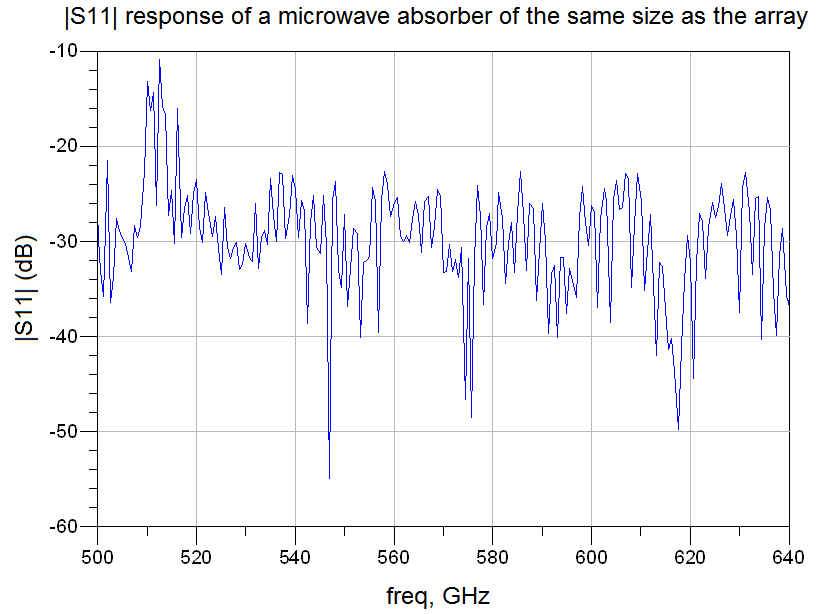}
   \end{tabular}
   \end{center}
   \caption[example] 
   { \label{fig:example} 
$S_{11}$ measurement of the absorber used to test the beam size. The absorption is roughly -30 dB which indicates that the beam spot is smaller than the rectangular piece. Note that there is a reflection peak at 515 GHz due to an error uncorrected by the calibration.}
   \end{figure} 
    
\subsection{Results}
The magnitude of the reflection coefficient $|S_{11}|$ of the bowtie array is shown in figures 8(a) and 8(b) corresponding respectively to the room temperature and the $70^{\circ}$ Celsius response. For both graphs, the curves in blue (solid line) represent the calibrated measured data while the curves in black (dotted line) represent the simulated result using Agilent's Advanced Design System (ADS). As one can see, there is a sharp absorption dip at $541$ GHz for the semiconducting state (figure 8(a)), which is due to the coupling of energy to the VO\textsubscript{2} film. As the temperature increases above the transition temperature (figure 8(b)), there is a large change in the response characterized by the absence of the absorption dip. This corresponds to the array being reflective (metallic state). The phase responses corresponding to the magnitude plots are displayed in figures 9(a) and 9(b). The circuit model is in good agreement with the measured data, although some measured magnitude values exceed 0 dB due to calibration error. We also note that there some discrepancies above 700 GHz due to multimode propagation not accounted for in the model.

\begin{figure}[ht!]
     \begin{center}
        \subfigure[]{%
            \label{fig:first}
            \includegraphics[width=0.46\textwidth]{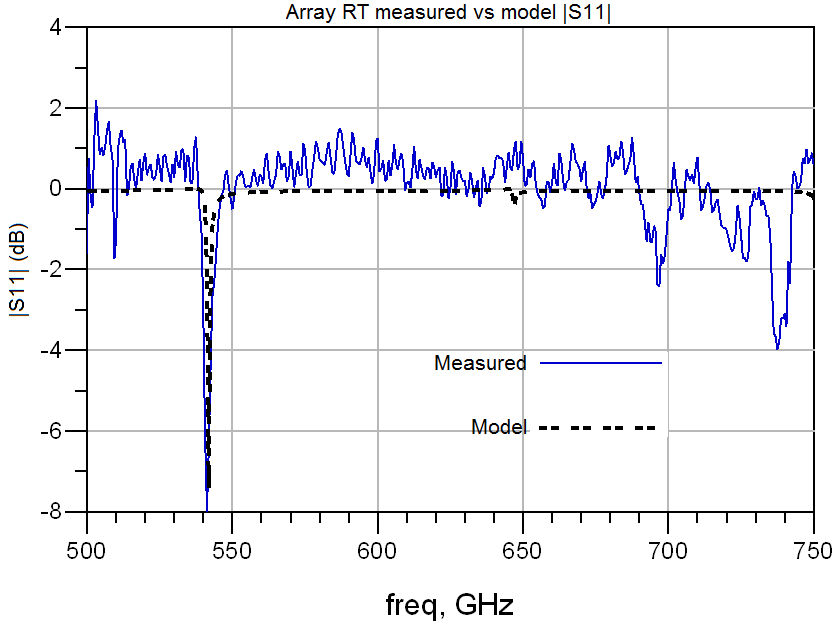}
        }%
        \subfigure[]{%
           \label{fig:second}
           \includegraphics[width=0.5\textwidth]{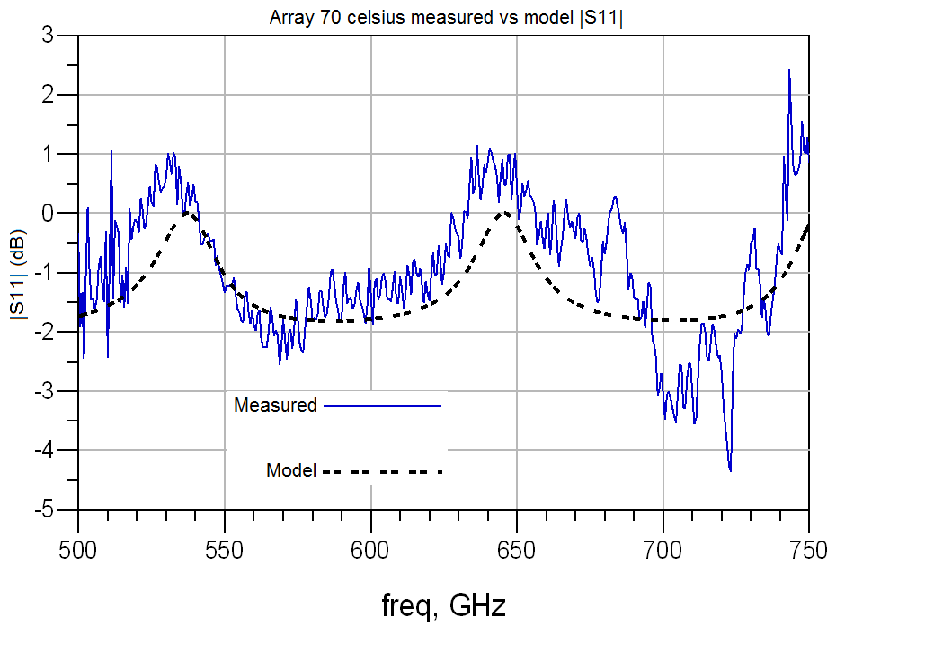}
        }\\ 
        
    \end{center}
    \caption{
        (a) The magnitude of $S_{11}$ at room temperature in dB (semiconducting state). The absorption dip at 541 GHz shows coupling of the beam to the $VO_2$ film. (b) The magnitude of $S_{11}$ at 70\degree{}C in degrees (metallic state). The absorption dip disappears which corresponds to the array being reflective.
     } 
     \label{fig:subfigures}
\end{figure}
\begin{figure}[ht!]
     \begin{center}
        \subfigure[]{%
            \label{fig:first}
            \includegraphics[width=0.44\textwidth]{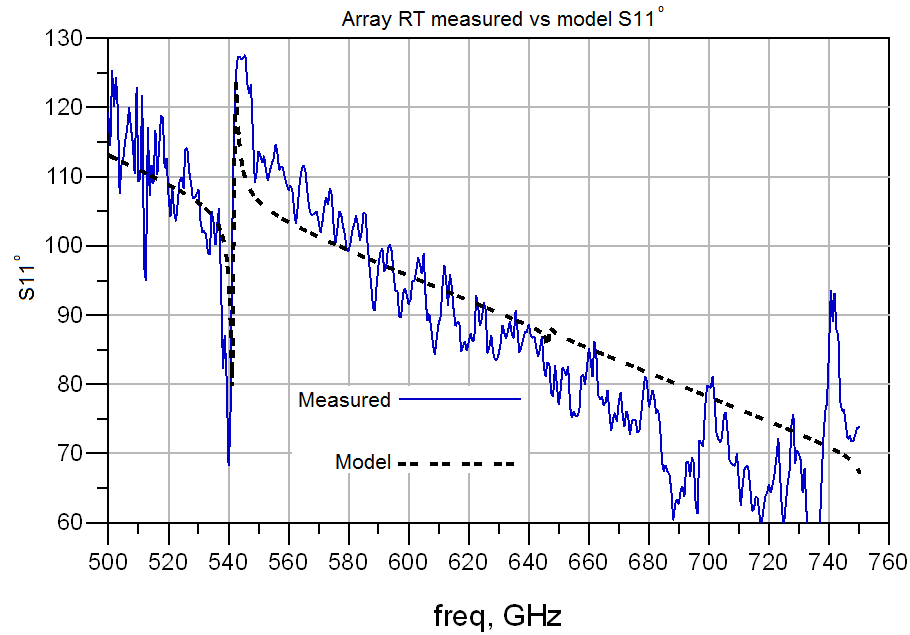}
        }%
        \subfigure[]{%
           \label{fig:second}
           \includegraphics[width=0.44\textwidth]{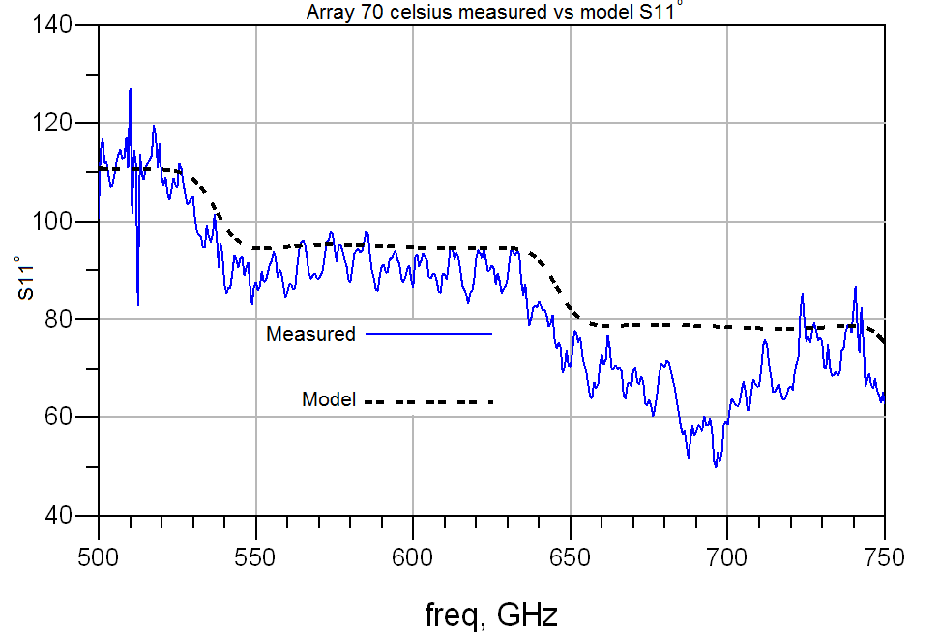}
        }\\ 
        
    \end{center}
    \caption{
        (a) The phase of $S_{11}$ at room temperature in degrees.(b) The phase of $S_{11}$ at 70\degree{}C in degrees.
     } 
     \label{fig:subfigures}
\end{figure}
\newpage
\section{Array Transmission simulation} \label{sec:sections}
Based on the model and the results from the previous section, the transmission response ($S_{21}$) of the mask, shown in figure 10, can be predicted. As we can see from figure 10(a), there is a change in transmission across the band when the array is heated. The maximum modulation is obtained at 660 GHz and corresponds to a 28 dB change in transmission between the two states. We do note however that in the metallic state, the mask is not completely transmissive (insertion loss of 12 dB). \\
\\

\begin{figure}[ht!]
     \begin{center}
        \subfigure[]{%
            \label{fig:first}
            \includegraphics[width=0.5\textwidth]{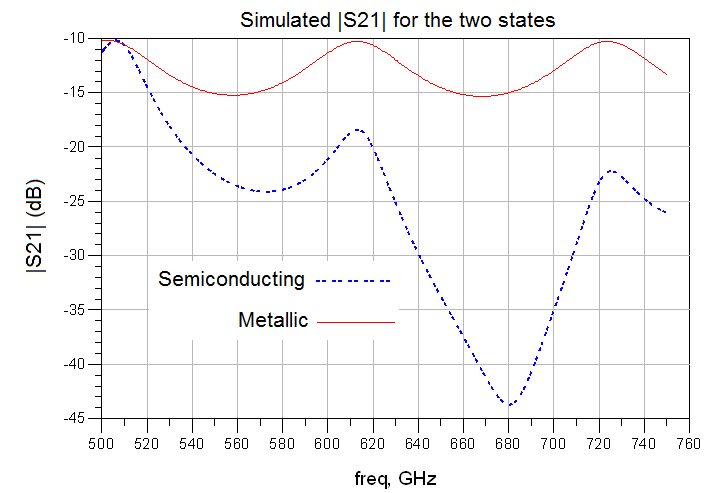}
        }%
        \subfigure[]{%
           \label{fig:second}
           \includegraphics[width=0.5\textwidth]{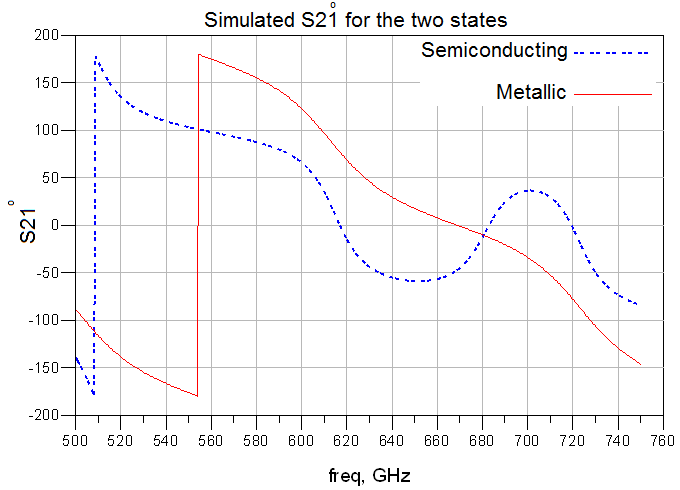}
        }\\ 
        
    \end{center}
    \caption{
        (a) The simulated magnitude of $S_{21}$ for the semiconducting and metallic state.(b) The simulated phase (in degrees) of $S_{21}$ for the semiconducting and metallic state. 
     } 
     \label{fig:subfigures}
\end{figure}

\section{Optically-Induced Switching} \label{sec:sections}
The transition of the VO\textsubscript{2} can also be triggered optically using laser illumination. An experiment was set up to demonstrate this. As the power of the laser used is 30 mW and focused on fraction of the array, only a small portion of the VO\textsubscript{2} bridges will switch and others will not. By increasing the temperature to $60^{\circ}$ Celsius, the array can be made to operate close to the switching transition. The reflection response at $60^{\circ}$ Celsius is shown in solid blue (figure 11). The response under illumination at the same temperature is shown as a dashed line. Finally we increase the temperature to $70^{\circ}$ Celsius switching the entire array (dotted red). As one can see, there is a change in $|S_{11}|$ of approximately 5 dB due to the laser. However, the optical flux in this experiment was not sufficient for a full transition to occur. 

\begin{figure}[h]
   \begin{center}
   \begin{tabular}{c}
   \includegraphics[height=6cm]{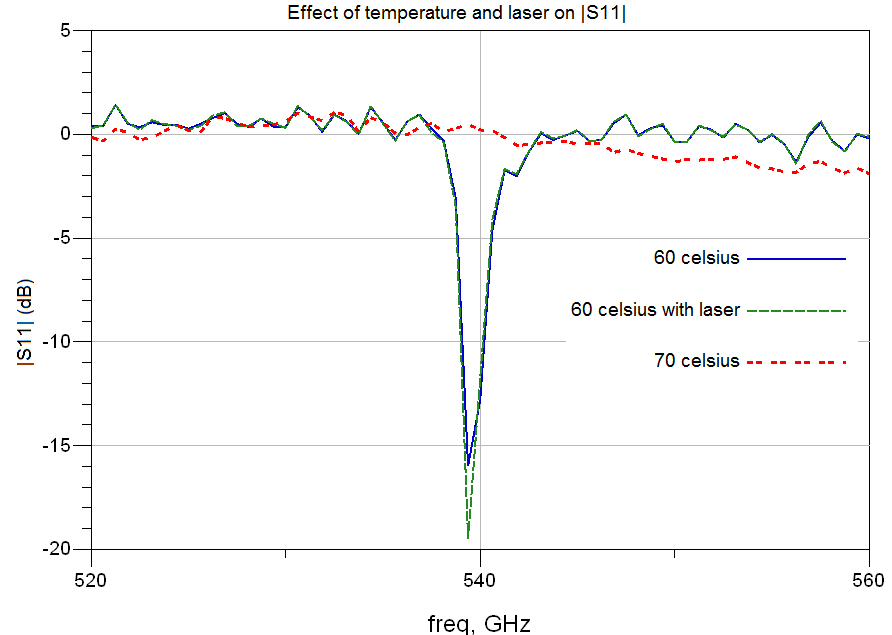}
   \end{tabular}
   \end{center}
   \caption[example] 
   { \label{fig:example} 
$|S_{11}|$ measurement of the array that shows the effect of the laser and temperature. Upon illumination with the laser, there is a change is reflection of approximately 5 dB at room temperature  . After increasing the temperature, we see that the array switches completely. Our conclusion is that the laser is not powerful enough and does not cover enough of an area to switch all the bowties.}
   \end{figure} 
   \newpage
\section{CONCLUSION}
\label{sec:conclusion} 
We have designed and characterized the reflection response of a bowtie antenna array coded aperture mask with VO\textsubscript{2} as the switching element. We have constructed a model for the array that is in good agreement with our measurement. Simulation of the transmission response of the mask shows that modulation of the transmission response between the two states is achieved over the WR-1.5 frequency band, with a maximum modulation of 28 dB at 660 GHz. We have also demonstrated the potential of switching the array optically using a laser.

\end{document}